\documentclass[conference, a4paper]{IEEEtran}
\IEEEoverridecommandlockouts
\usepackage{cite}
\usepackage{tfrupee}
\usepackage{amsmath,amssymb,amsfonts}
\usepackage{algorithmic}
\usepackage{graphicx}
\usepackage{textcomp}
\usepackage{xcolor}
\usepackage{algorithm, algorithmic}
\usepackage{array}
\usepackage{url}
\usepackage[caption=false,font=footnotesize]{subfig}
\def\BibTeX{{\rm B\kern-.05em{\sc i\kern-.025em b}\kern-.08em
    T\kern-.1667em\lower.7ex\hbox{E}\kern-.125emX}}

\IEEEoverridecommandlockouts\IEEEpubid{\makebox[\columnwidth]{ 978-1-7281-2472-8/19/\$31.00~\copyright~2019 IEEE \hfill} \hspace{\columnsep}\makebox[\columnwidth]{ }}
\begin{document}

\title{Realization of Self--Demand Response Through Non--Intrusive Load Monitoring Algorithm}

\author{\IEEEauthorblockN{Shashank Singh}
\IEEEauthorblockA{\textit{Electrical \& Electronics Engineering}
\\\textit{NIT Tiruchirappalli}
\\Tiruchirappalli, India
\\shashanksingh0110@gmail.com}
\and
\IEEEauthorblockN{Sudharsan Thirumalai}
\IEEEauthorblockA{\textit{Electrical \& Electronics Engineering}
\\\textit{NIT Tiruchirappalli}
\\Tiruchirappalli, India
\\stsudharshan@gmail.com}
\and
\IEEEauthorblockN{M.~P.~Selvan}
\IEEEauthorblockA{\textit{Electrical \& Electronics Engineering}
\\\textit{NIT Tiruchirappalli}
\\Tiruchirappalli, India
\\selvanmp@nitt.edu}
\thanks{This publication is an outcome of the R\&D work undertaken in the project under the Visvesvaraya PhD Scheme of Ministry of Electronics and Information Technology, Government of India, being implemented by Digital India Corporation (formerly Media Lab Asia).}
}

\maketitle

\begin{abstract}
The real--time statistics on key consumer parameters and key utility parameters earmark the implementation of demand response (DR) under the smart grid (SG) paradigm. Advanced metering infrastructure (AMI) enables monitoring and control over both key parameters to reflect reliable information on SG. Firstly, this paper aims at a physical realization of an AMI, which comprises a meter data management system (MDMS) supported by smart meters. Secondly, a sliding window based non--intrusive load monitoring algorithm is proposed to illustrate the power consumption pattern. The MDMS features an incremental block--rate tariff structure given by Tamil Nadu state electricity board for the estimation and prediction of electricity bill. Finally, an implementation perspective is presented describing the features of its constituents both theoretically and experimentally, which could envision a consumer--facing grid and exert the proposed self--DR scheme.
\end{abstract}
\renewcommand\IEEEkeywordsname{Keywords}

\section{Introduction}
The modus operandi of electrical power grids is experiencing a renaissance. The integration of highly reliable, efficient, security--driven information and communication technology (ICT), and advanced data analytics are transforming traditional electrical power grids into smart grids (SG). The modern power system or the SG is a heterogeneous system, where the interaction among its digital elements, physical elements, consumers, and the environment happens in a closed manner. Internet of things (IoT) is a buzzword of this decade, that refers to the interconnection of machines and devices with internet. The convergence of IoT with traditional power grids is transcending the limitations of the power grids and propelling it beyond singularity by bridging the gap between the consumers and the grids \cite{collier2017}.

Demand response (DR) is one of the key provisions present under SG environment, which plays a major role in healthy operation of the grid. DR schemes provide chances to the consumers to reduce their peak load demand in response to some form of financial incentives or demand based tariffs \cite{palensky2011}, thereby increase the consumers' participation.

\begin{figure}[t]
\centerline{\includegraphics[width = 0.7\columnwidth]{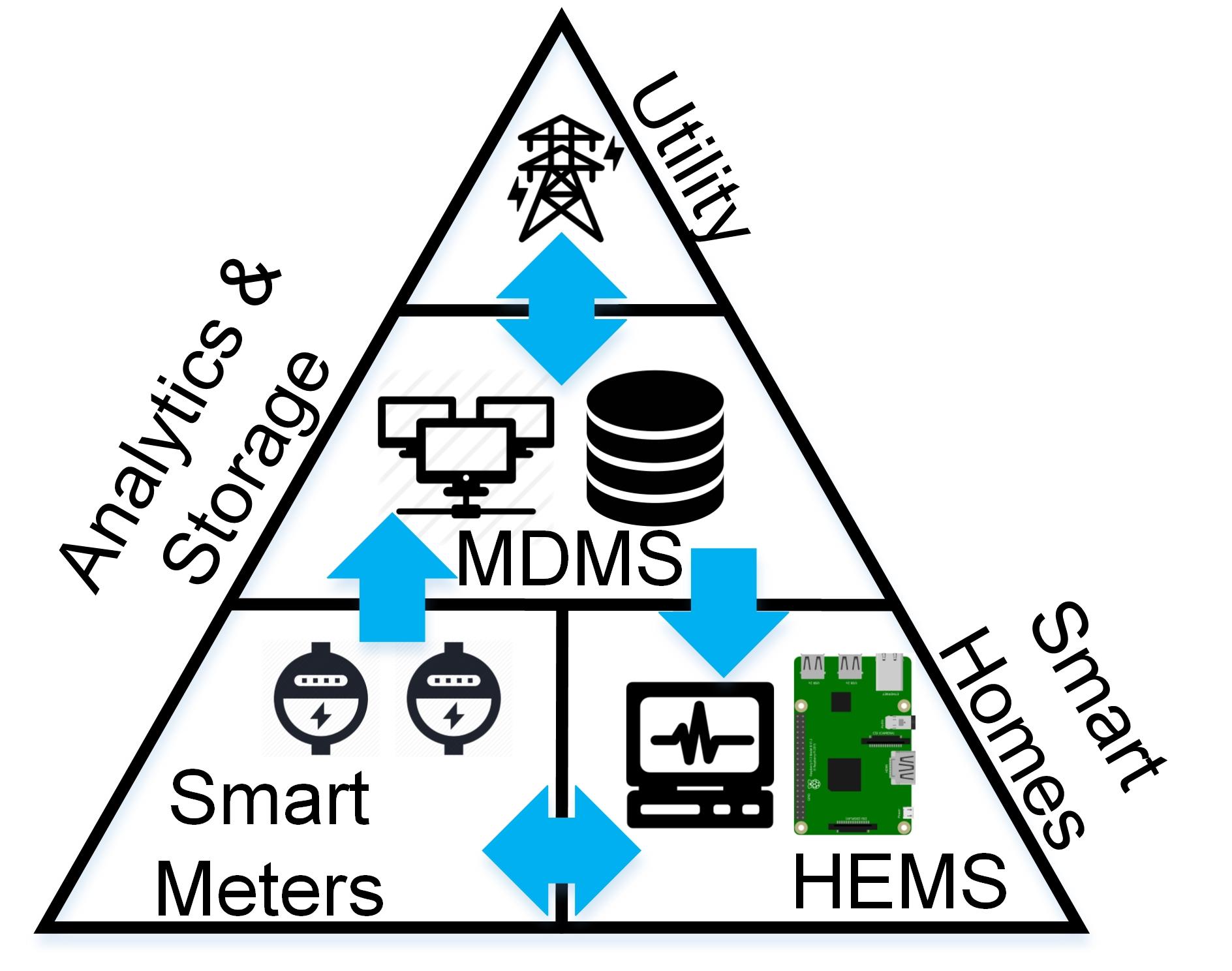}}
\caption{Hierarchical structure of an AMI}
\label{AMI}
\end{figure}

Advanced metering infrastructure (AMI) is an integrated system of smart meters (SM) \cite{singh2017}, communication networks, and meter data management systems (MDMS) that enables bidirectional communication between utilities and consumers. The architecture of AMI is illustrated in Fig.~\ref{AMI}, which emphasizes SM(s) or home energy management system \cite{singh2018} as an essential element deployed at the consumer side.  Reference \cite{giaconi2018} suggests that SM(s) in the United Kingdom and Texas, USA send measurements at every 30 and 15 minutes respectively. However, there are challenges in the pathway of deployment, application and data analysis of SM(s). Thus, the rapid development of various DR algorithms and large--scale integration of DR schemes could require the frequent transmission of measurements.  

Furthermore, the real--time load monitoring (LM) under AMI has several benefits. LM acknowledges consumers on their daily consumption pattern, which could encourage them to participate in DR programs. On top of that,  the utilities can also optimally schedule their energy resources, and possibly use data mining techniques to forecast the demand. LM has two categories namely \textit{intrusive} and \textit{non--intrusive}. Intrusive LM requires equipping a dedicated sensor to each household appliances, however non--intrusive load monitoring (NILM) discovers the contribution of individual appliances from aggregate power signature obtained via single SM. The development of NILM techniques lies back in the 80s, where it was originally developed in MIT, USA. Reference \cite{zeifman2011} reviews the continuous developments in NILM techniques and reveals that there are no generalized NILM techniques for all appliances, and the developments are still ongoing.  

In \cite{molina2010}, researchers extracted activity and information of consumers such as the number of people in a house, their eating and sleeping patterns from the aggregate load profile, and proposed a need of privacy--centric architecture for SM(s). A prototype of a direct load control scheme using NILM was built and tested in \cite{bergman2011}, wherein the utility benefits the consumers if they agree to provide control onto their registered loads.  Reference \cite{shaloudegi2016} proposed a Markov model based NILM algorithm, which yielded 50\% better precision than the references therein. A graph mapped the active power dataset of a consumer, which formulated a basis of graph--based NILM in \cite{kumar2017}. The short term load forecasting using the results of NILM along with artificial neural network was detailed in \cite{ebrahim2018}. Additionally, authors in \cite{marchiori2011, lu2017, Liu2019} presented the continuous advancements in NILM methods.

The aforementioned existing literature suggests diverse approaches developed by the researchers, wherein many of the NILM methods estimate the run time of appliances using the collected set of measurements from SM, whereas few methods are based on heuristics techniques, and supervised learning methods by incorporating past loads signatures.

Unlike the previous works, this paper showcases a new implementation perspective on AMI, wherein the MDMS layer hosts a sliding window based NILM algorithm. This paper also describes the software level development of MDMS. Herein, the MDMS receives data from consumers via SM and execute NILM. Finally, the concept of self--DR, which was proposed in \cite{singh-SM2019}, is improvised by integrating the analysis of NILM. Hence, the major developments presented in this paper are as follows:
\begin{itemize}
\item Design of MDMS using open--source software tools, wherein MDMS provides a web--based user interface to the consumer.
\item Development and testing of NILM algorithm with real loads.
\item Depiction of insight of the consumption pattern of the consumers on MDMS using described NILM algorithm.
\item Improvised Self--DR technique for consumers. 
\end{itemize}

The paper is organized in the following way. Section II deals with the development of MDMS and introduces self--DR. Section III presents the formulation of NILM algorithm and its performance analysis. Finally, section IV concludes the paper.

\section{Meter Data Management System and Self--Demand Response}
\begin{figure}[htbp]
\centering
\includegraphics[width = 0.8\columnwidth]{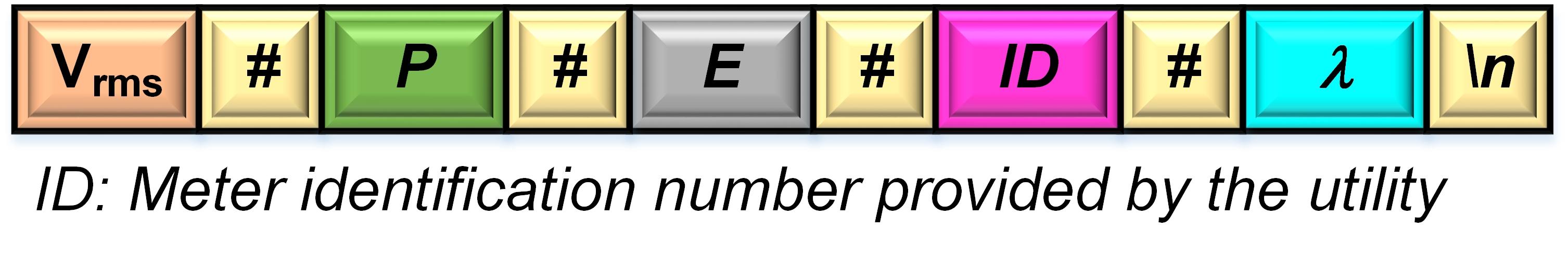}
\caption{Adapted data format for communication between SM and MDMS}
\label{DataFormat}
\end{figure}

\begin{figure}[htbp]
\includegraphics[width= \columnwidth]{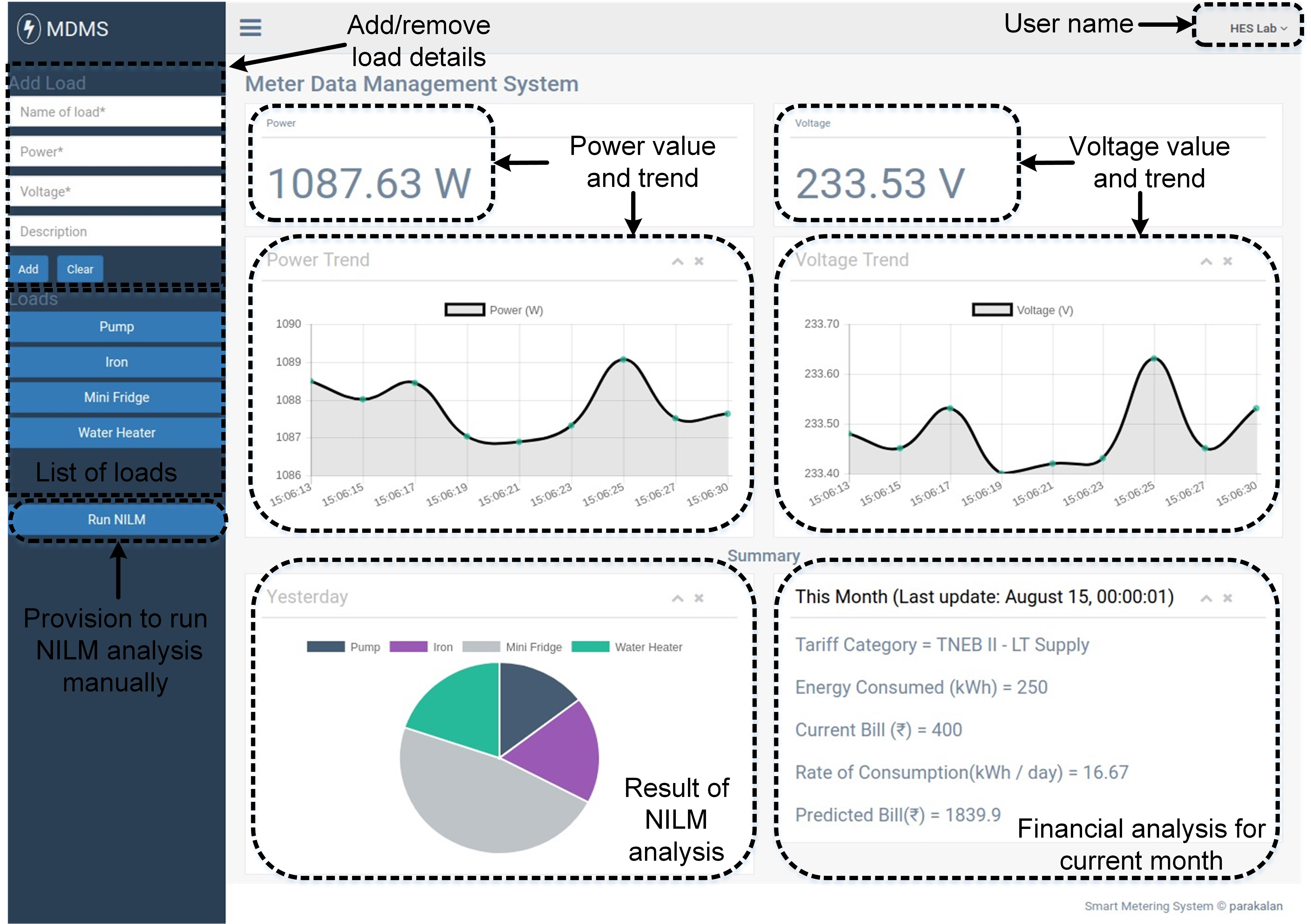}%
\caption{A web app based user interface of MDMS}
\label{User Interface}
\end{figure}

Under the paradigm of AMI, the SM typically transmits a consumer's consumption pattern to the utility's server. The development of a SM used for data acquisition and exchange is outlined in \cite{singh-SM2019}, wherein the SM follows a hash--separated format (Fig.~\ref{DataFormat}) to formulate a data packet. The MDMS retrieves that data frame and executes a computer program to perform data analytics and storage tasks. Therefore, a web app has been developed in the laboratory using high--level Python based web framework \verb"Django" , which showcases the operations of MDMS. This web app contains a computer program, which has two components: front--end and back--end. The data frame, which traverses through a communication network and reaches the MDMS. Thereafter, a back--end program at MDMS splits data using $\#$ as format specifier and stores the values in \verb"MySQL" database. The back--end program also analyzes the retrieved data and executes NILM algorithm. The front--end component presents the web--based user interface (Fig.~\ref{User Interface}), which displays the trends, and results from NILM analysis.

\subsection{Description of the Database Model}
\begin{figure}[htbp]
\includegraphics[width=\columnwidth]{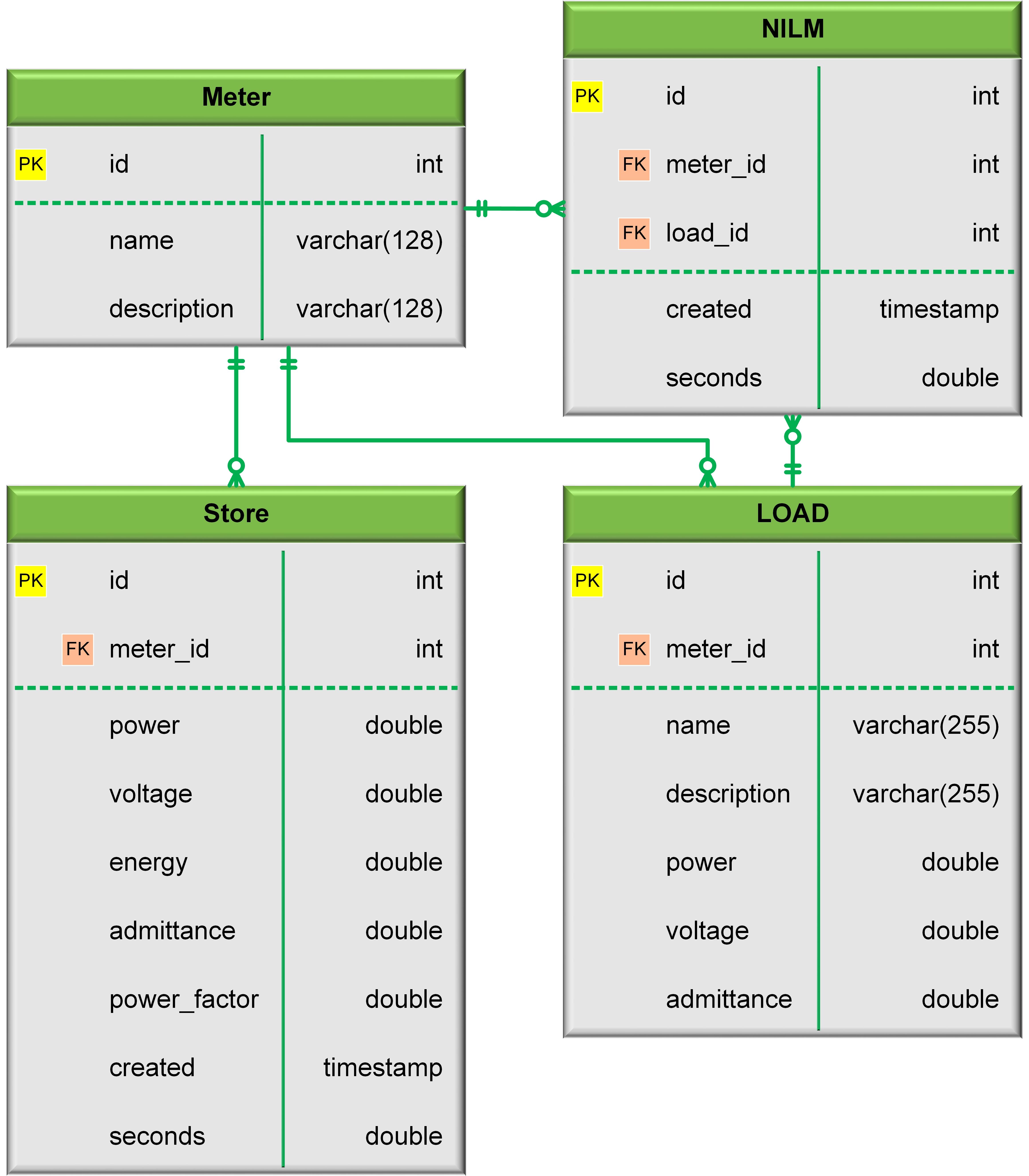}%
\caption{Depiction of MDMS database schema using crow's foot entity relationship diagram}
\label{DB Schema}
\end{figure}

The schema of database is shown in Fig.~\ref{DB Schema}, which contains four tables namely: \textit{meter}, \textit{NILM}, \textit{store}, and \textit{load}. The \textit{meter} table stores the unique meter identification number, which is provided by the utility to their consumers. The \textit{load} table stores the electrical specifications of every registered appliance, e.g., rated voltage, power and power factor. The \textit{store} table stores the real--time data received from the meter. The \textit{NILM} table stores the result of NILM analysis, which has the attributes namely: run time of an appliance, its name (id) and node id (meter id). The type of relationship between these tables in the database is \textit{one--to--many} among \textit{meter} and \textit{load}, \textit{meter} and \textit{store}, \textit{meter} and \textit{NILM}, \textit{load} and \textit{NILM}.

A data visualization library known as \verb"Chart.js" fetches data from \textit{store} table and renders the power and voltage trends on front--end of the web app. At the end of a day, the back--end program fetches the stored values from \textit{store} table of the database to perform following tasks: NILM analysis, estimation of the rate of power consumption till the current date of that month, and prediction of the bill for that month. Thereafter, the front--end program fetches the results of NILM analysis from the \textit{NILM} table and displays it on the user panel of the web app through \verb"Chart.js" library.

\subsection{Self--Demand Response}
The Southern California Edison Power Station initiated the reduction of electricity demand by introducing \verb"Energy Orb" \cite{chapa2007}, a glowing sphere that changes its color indicating the variations in electricity price. It reportedly brought a $40\%$ reduction in energy usage for consumers. It was a non--intrusive approach persuading consumers to change their consumption patterns according to the color of \verb"Orb". There are other similar devices introduced in the past \cite{chapa2007}, \cite{rich2006}. Furthermore, another non--verbal approach named \verb"ColorPower" was proposed by an MIT startup, which gives color based indications to consumers alerting them to exert control on power consumption during peak hours \cite{stauffer2012} \cite{ranade2010}. 

The idea of self--DR, inspired from the initiatives above, is defined as the actions by consumers to alter their consumption pattern, without explicitly being asked or forced by the utility, by responding psychologically to variations in electricity price and predicted data.

Furthermore, the presented work incorporates an incremental block rate tariff (IBRT), which combines the results from NILM analysis and showcases the idea of self--DR. The IBRT structure is followed by Tamil Nadu state (in India) electricity board for its low tension consumers \cite{tneb2017}.  The daily electricity bill and rate of power consumption predict monthly electricity bill using IBRT.  Upon combining predicted results with the result of NILM analysis would encourage consumers to alter their consumption patterns in order to reduce the monthly electricity bill. Such response of a consumer is referred to as self--DR, a feature of this system, which reflects the changes in consumer's consumption pattern.

\section{The NILM Algorithm}
The presented NILM algorithm requires electrical specifications of appliances. The MDMS enables consumers to input the specifications of the intended appliances via web app to participate in self--DR scheme. The proposed sliding window--based NILM algorithm (SWNA) computes the share of intended/registered appliances in the total energy consumption for a particular day. The MDMS has been configured to execute SWNA at the end of every day automatically. Further, it also provides an option to execute SWNA on consumer's request.

\subsection{Sliding Window Based NILM Algorithm (SWNA)}
Assuming the binary operating status of $i^{th}$ appliance at an instant $t$ is $flag^i_t$, active power consumption $P^i_t$, and the total number of participating appliances is $\alpha$, then the aggregated power consumption $(P_t)$ of a house at an instant $t$ can be represented by \eqref{eq10}. In a practical scenario, the net power consumed by an appliance at any instant shall be unequal to its rated value, but it could be close to its rating given the fluctuations in the supply voltage. Hence, a term $pwr\_tol$ has been introduced to represent the percentage tolerance in the rated power consumption of an appliance due to grid conditions. Thereby, the boundary conditions for \eqref{eq10} can be represented by \eqref{boundary}.
\begin{equation} 
P_t= \sum_{i=1}^{\alpha}{\left(flag_t^i \times P_t^i \right)}
\label{eq10}
\end{equation}
\begin{equation}
\sum_{i=1}^{\alpha}{\left(flag_t^i \times P_t^i \times lval_t^i \right)} < P_t < \sum_{i=1}^{\alpha}{\left(flag_t^i \times P_t^i \times rval_t^i \right)}
\label{boundary}
\end{equation}
where, $pwr\_tol$ is assumed to be $10\%$ in this study, \\
$lval=(1-pwr\_tol)$, and 
$rval=(1+pwr\_tol)$
\begin{figure}[htbp]
\centering
\subfloat[Scanning of time--varying power data using sliding window]{\includegraphics[width = \columnwidth]{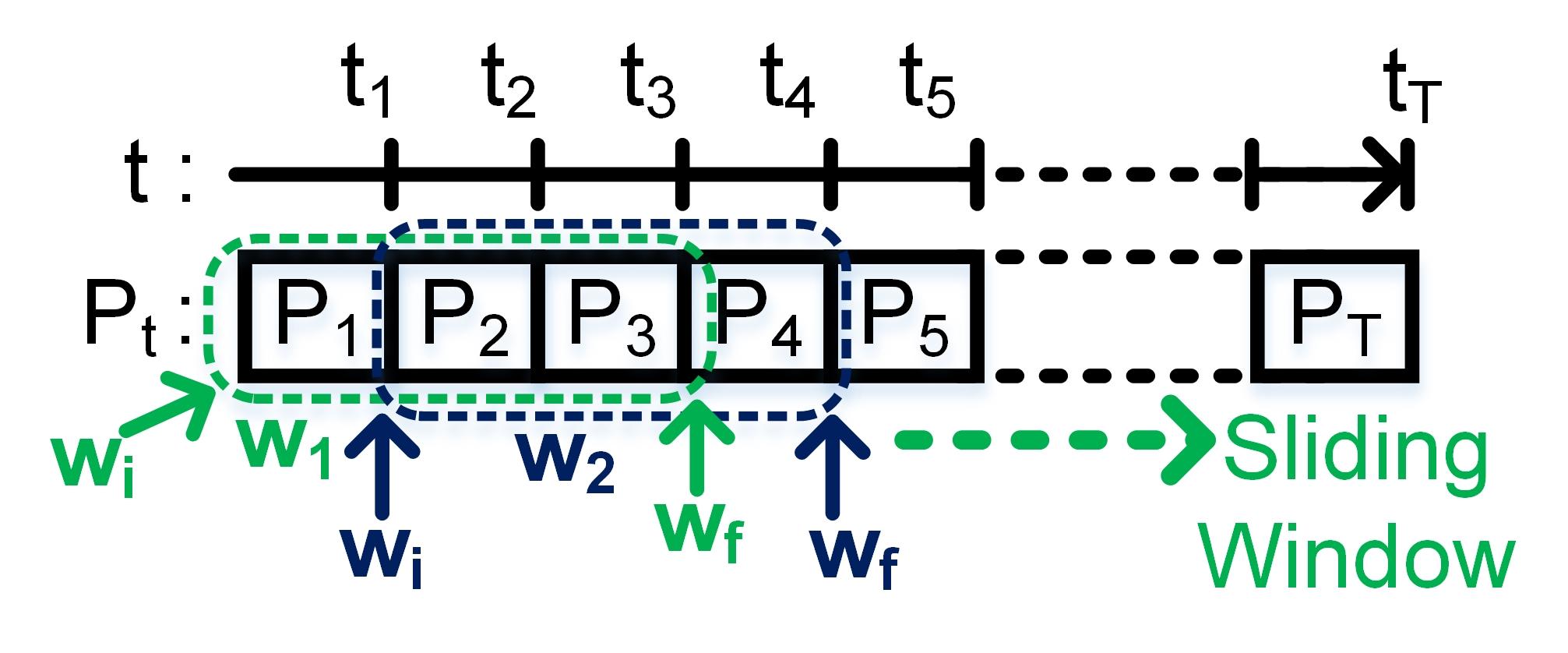}\label{NILMScanning}}
\hfill
\subfloat[Detection of transient values during load switching]{\includegraphics[width=\columnwidth]{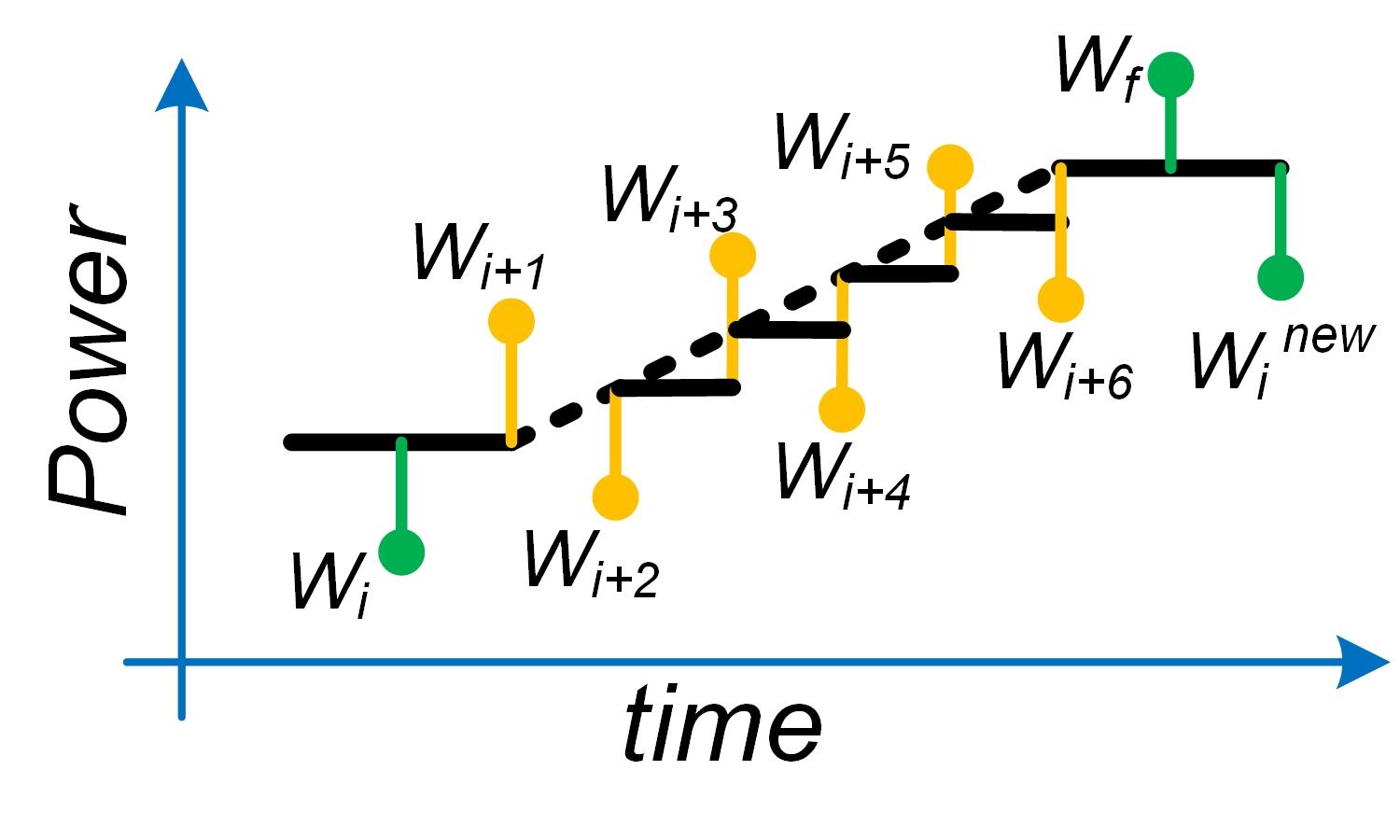}\label{NILMTransient}}
\caption{Depiction of sliding window based NILM algorithms}
\label{NILMSliding}
\end{figure}

Initially, the algorithm fetches the power and corresponding time--stamp data from the \textit{Store} table, and scans using a window having its default length ($win\_len$) equal to three as shown in Fig.~\ref{NILMScanning}. Parameters $w_i$ (initial instant, represents steady state value of $P_t$, e.g., $P[w_i]=P_1$) and $w_f$ (final instant, represents steady state value of $P_t$) define the lower and upper boundaries of the window, where an instant $w_{i+1}$ reflects transition period and the power consumption corresponds to that instant $P[w_{i+1}]$ represents intermediate power level between $P[w_i]$ to $P[w_f]$. 

However, there is a possibility to have multiple intermediate power levels between $w_f$ and $w_i$ as shown in Fig.~\ref{NILMTransient}: an appliance could take more than default $win\_len$ instants to reach to its steady state power consumption. Under such conditions adhering to default value of $win\_len$ could be bottleneck for this algorithm. Hence, this algorithm incorporates another novel feature to track intermediate transition periods by dynamically expanding the length of its search window. Transition tolerance has been introduced to incorporate the detection of intermediate transition periods. \verb"TRANS_TOL" (considered as $5W$ in this study) represents the apparent small difference in power levels obtained from \eqref{eq10} after two successive queries. The \verb"TRANS_TOL" reflects the fluctuations in rated power consumption during ON condition of an appliance, which should not be interpreted as major transitions in power levels. 

After considering aforementioned parameters, and initial conditions: $win\_len \leftarrow 3$ (default value), $flag \gets 0$, the following major states of an appliance can be detected from Fig.~\ref{NILMSliding}:\\
\textbf{Steady State}:\\ $if$ $|P[w_{f+1}]-P[w_f]| \leq$ \verb"TRANS_TOL" \\
\textbf{Load ON}:\\ $if$  $lval <|P[w_i]-P[w_f]| < rval$ $\&$ $flag == 0$ \\
\textbf{Load OFF}:\\ $if$  $lval <|P[w_i]-P[w_f]| < rval$ $\&$ $flag == 1$ \\
\textbf{Transition $(\delta)$}:\\ $if$ $|P[w_{f+1}]-P[w_f]| >$ \verb"TRANS_TOL"

If there was only one instant representing $\delta$ then the upper bound of the window would be $w_f = w_i + win\_len - 1$. However, in the event of the occurrence of multiple instants representing $\delta$, then the algorithm keep tracking the subsequent occurrences of $\delta$, each followed by a dynamic increment in $win\_len$ until the steady state is detected. Thereafter, the $win\_len$ is initialized to its default value followed by $w_i^{new} \leftarrow w_{f+1}$ for subsequent tracking. The developed and implemented SWNA using \textit{object oriented programming} concept has been explained using a \verb"pseudocode".  
\begin{algorithm}[!htbp]
\caption{Pseudocode of NILM Algorithm}
\begin{algorithmic}[1]
\STATE \textbf{Class}~\verb"Loads":
\STATE ~\textbf{Function}~\verb"init"
\STATE ~~$v \leftarrow 230$		\COMMENT{Indian Standard}
\STATE ~~$name \leftarrow name$ \COMMENT{Name of the load}
\STATE ~~$pwr \leftarrow pwr$ 	\COMMENT{Rated power}
\STATE ~~$t_{start} \leftarrow 0.0$   \COMMENT{Load start time}
\STATE ~~$t_{stop} \leftarrow 0.0$   \COMMENT{Load stop time} 
\STATE ~~$flag \leftarrow 0$    \COMMENT{status flag}
\STATE ~~$dur \leftarrow 0.0$   \COMMENT{duration of operation}
\STATE ~\textbf{End Function}
\STATE \textbf{End Class}
\STATE \verb"Loads" \COMMENT{List of all loads}
\STATE \verb"PWR" \COMMENT{Power consumption at different time instances}
\STATE \verb"TIME" \COMMENT{time instances}
\STATE \verb"PWR_TOL" \COMMENT{Power Tolerance}
\STATE \verb"TRANS_TOL" \COMMENT{Transition Tolerance}
\STATE \textbf{Function} \verb"NILM(PWR, PWR_TOL)" 
\STATE ~$i$ $\leftarrow$ $0$
\STATE ~\textbf{While} $i~<$ \textit{len~of}~(\verb"PWR") 
\STATE ~~$win\_len$ $\leftarrow$ $2$
\STATE ~~\textbf{For} $j = 1$ $\rightarrow$ \textit{len~of}~(\verb"LOADS")
\STATE ~~~$lval$ $\leftarrow$ \verb"Loads"$[j]$.\verb"pwr"$\times$(1$-$\verb"PWR_TOL")
\STATE ~~~$rval$ $\leftarrow$ \verb"Loads"$[j]$.\verb"pwr"$\times$(1$+$\verb"PWR_TOL")
\STATE ~~~\textbf{While} ($i+win\_len$) $<$ \textit{len~of}~(\verb"PWR")
\STATE ~~~~$change$ $\leftarrow$ \verb"PWR"$[i+win\_len]-$\verb"PWR"$[i]$
\STATE ~~~~\textbf{If} $lval~<~|change|~<~rval$
\STATE ~~~~~$x \leftarrow i+win\_len$
\STATE ~~~~~\textbf{If} \verb"|PWR"$[x]-$\verb"PWR"$[x+1]|<$ \verb"TRANS_TOL"
\STATE ~~~~~~\textbf{If} \verb"LOADS"$[j]$.$flag$ is 1
\STATE ~~~~~~~\verb"Loads"$[j].t_{stop}\leftarrow$\verb"TIME"$[i]$
\STATE ~~~~~~~\verb"Loads"$[j].flag\leftarrow$~0
\STATE ~~~~~~~$y \leftarrow$ \verb"Loads"$[j].t_{stop}-$\verb"Loads"$[j].t_{start}$
\STATE ~~~~~~~\verb"Loads"$[j].dur \leftarrow$ \verb"Loads"$[j].dur+y$ 
\STATE ~~~~~~\textbf{Else}
\STATE ~~~~~~~\verb"Loads"$[j].t_{start} \leftarrow $\verb"TIME"$[i]$
\STATE ~~~~~~~\verb"Loads"$[j].flag\leftarrow$~1
\STATE ~~~~~~\textbf{EndIf}
\STATE ~~~~~~$i \leftarrow i+win\_len$
\STATE ~~~~~~$win\_len \leftarrow 2$
\STATE ~~~~~~$break$
\STATE ~~~~~\textbf{Else}
\STATE ~~~~~~~$win\_len \leftarrow win\_len+1$
\STATE ~~~~~\textbf{EndIf}
\STATE ~~~~\textbf{EndIf}
\STATE ~~~\textbf{EndWhile}
\STATE ~~\textbf{EndFor}
\STATE ~\textbf{EndWhile}
\STATE \textbf{EndFunction}
\STATE \textbf{Call} \verb"NILM(PWR, PWR_TOL)" 
\end{algorithmic} 
\end{algorithm}

\subsection{Performance Analysis of NILM Algorithm}
\begin{figure}[htbp]
\centering
\includegraphics[width=\columnwidth]{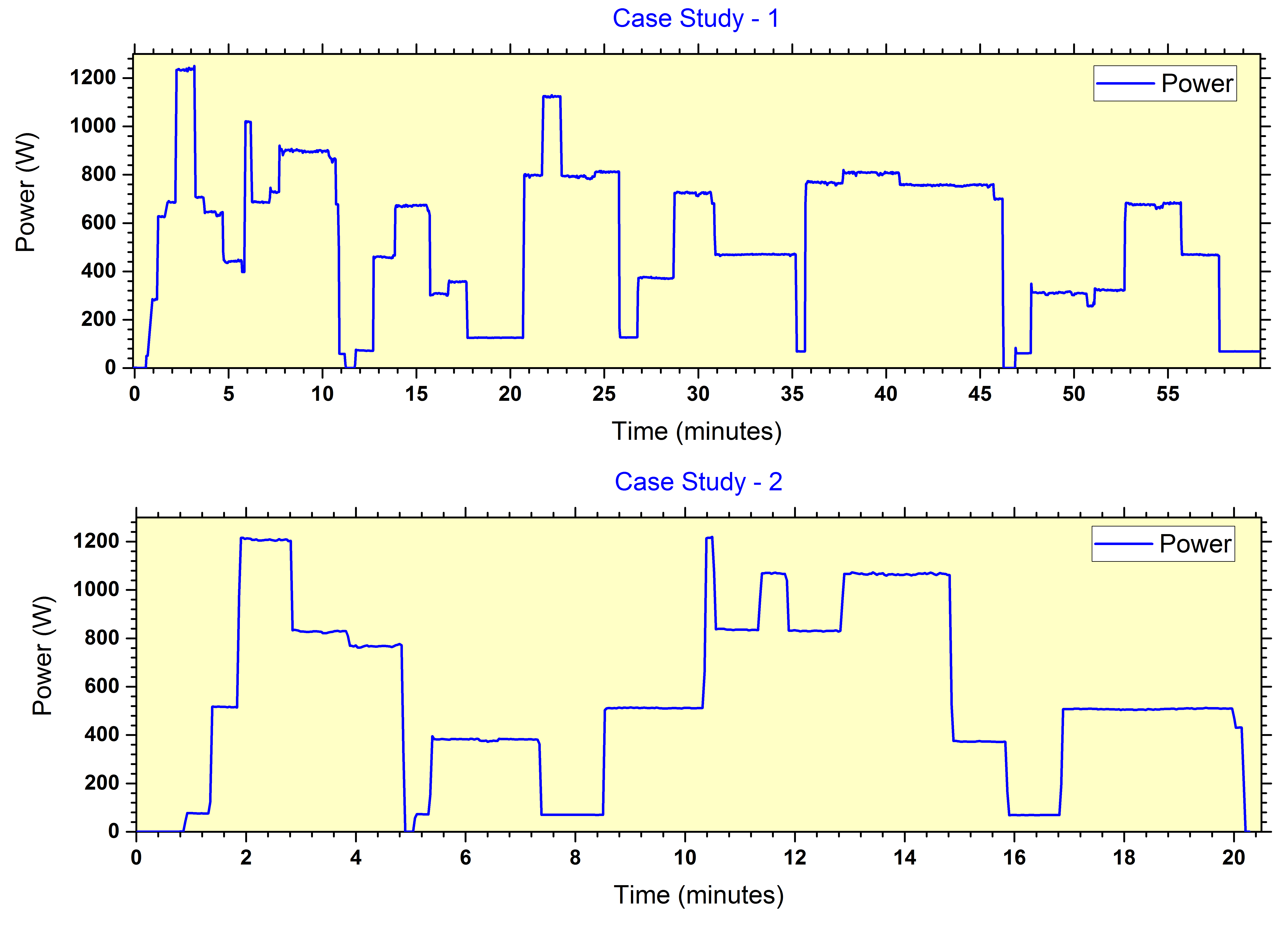}
\caption{Load curves used to analyze the performance of SWNA}
\label{CurveNILM}
\end{figure}

This section emphasizes the two case studies, which are conducted to test the accuracy of SWNA. Table~\ref{loadsSWNA} describes the details of the appliances used in case studies. The duration of first and second case studies was 60 and 20 minutes respectively, wherein the appliances were switched randomly multiple times. The load curves obtained from these two case studies are illustrated in Fig.~\ref{CurveNILM}, which depicts the random switching of appliances during the intended duration of the case study.  The tabulated results (Tables~\ref{SWNA-1} and \ref{SWNA-2}) shows the performance of the NILM algorithm, which consists of the actual duration of operation of an appliance and the duration of operation obtained from SWNA. Tables~\ref{SWNA-1} and \ref{SWNA-2} imply that the error in computing the duration of the operation using SWNA was less than $1.5\%$.
\begin{table}[htbp]
\renewcommand{\arraystretch}{1.1}
\caption{Description of Loads used for Testing SWNA}
\label{loadsSWNA}
\centering
\begin{tabular}{c c c}
\hline
\hline
\textbf{S.No.} & \textbf{Name} & \textbf{Rating} \\
\hline
1 & Pump 			& 230V, 250W \\
2 & Mini Fridge		& 230V, ~70W \\
3 & Iron 			& 230V, 400W \\
4 & Water Heater	& 230V, 700W \\
\hline
\hline
\end{tabular}
\end{table}
\begin{table}[htbp]
\renewcommand{\arraystretch}{1}
\caption{Case Study 1: Duration 60 minutes}
\label{SWNA-1}
\centering
\begin{tabular}{c c c c c}
\hline
\hline
\textbf{S.No.} & \textbf{Name} & $D_{act}~(s)$ & $D_{SWNA}~(s)$ & $\epsilon$ \\
\hline
1 & Pump 			& 1365 & 1375 & 0.73\\
2 & Mini Fridge		& 2740 & 2741 & 0.03\\
3 & Iron 			& 1230 & 1231 & 0.08\\
4 & Water Heater	& 1295 & 1302 & 0.54\\
\hline
\hline
\multicolumn{5}{l}{$D_{act}~(s)$: Actual duration of operation (in seconds)}\\
\multicolumn{5}{l}{$D_{SWNA}~(s)$: Duration of operation (in seconds) obtained}\\
\multicolumn{5}{l}{from SWNA}\\
\multicolumn{5}{l}{$\epsilon$: Error in percentage}\\
\end{tabular}
\end{table}
\begin{table}[htbp]
\renewcommand{\arraystretch}{1.1}
\caption{Case Study 2: Duration 20 minutes}
\label{SWNA-2}
\centering
\begin{tabular}{c c c c c}
\hline
\hline
\textbf{S.No.} & \textbf{Name} & $D_{act}~(s)$ & $D_{SWNA}~(s)$ & $\epsilon$ \\
\hline
1 & Pump 			& 330 & 330 & 0.00\\
2 & Mini Fridge		& 970 & 981 & 1.13\\
3 & Iron 			& 435 & 429 & 1.37\\
4 & Water Heater	& 450 & 450 & 0.00\\
\hline
\hline
\end{tabular}
\end{table}

The satisfactory performance of SWNA depends upon the stiffness of the supply voltage and the accuracy of user--provided load parameters, but does not accept multiple loads of similar ratings. SWNA provides flexibility to the consumers by allowing them to choose intended appliances for monitoring. The outcome of SWNA emphasizes the dominance of intended appliances in total energy consumption as shown in the pie chart of Fig.~\ref{User Interface}. It is expected from the developed MDMS that the preceded NILM analysis will deliberately have a benign effect upon the consumers to adapt for self--DR.

\section{Conclusion}
Implementation of advanced metering infrastructure (AMI) has been demonstrated in this paper with a meter data management system (MDMS) providing an interactive environment for the consumers. Diverse data visualization methods, non-invasive, and persuasive interfaces allow households to explore their consumption pattern simply and straightforwardly. 

The key elements of AMI are MDMS, a non--intrusive load monitoring (NILM) and price prediction. Aforementioned key elements are practically implemented in the laboratory using open--source software frameworks. The NILM yields consumption pattern, which adds up to AMI's ability to implement demand response by applying a flexible tariff structure and providing appliance specific usage analytic. The concept of self--demand response has been discussed, which is based on the idea that the consumption analytic will influence consumption pattern to bring positive behavioral changes in energy consumption towards flattening the load curve. This developed prototype of AMI integrated with NILM holds great value for future development in load prediction and management algorithms, meter data management systems, and remote monitoring under the paradigm of Smart Grids.

\bibliographystyle{IEEEtran}
\bibliography{Ref}

\begin{thebibliography}{10}
\providecommand{\url}[1]{#1}
\csname url@samestyle\endcsname
\providecommand{\newblock}{\relax}
\providecommand{\bibinfo}[2]{#2}
\providecommand{\BIBentrySTDinterwordspacing}{\spaceskip=0pt\relax}
\providecommand{\BIBentryALTinterwordstretchfactor}{4}
\providecommand{\BIBentryALTinterwordspacing}{\spaceskip=\fontdimen2\font plus
\BIBentryALTinterwordstretchfactor\fontdimen3\font minus
  \fontdimen4\font\relax}
\providecommand{\BIBforeignlanguage}[2]{{%
\expandafter\ifx\csname l@#1\endcsname\relax
\typeout{** WARNING: IEEEtran.bst: No hyphenation pattern has been}%
\typeout{** loaded for the language `#1'. Using the pattern for}%
\typeout{** the default language instead.}%
\else
\language=\csname l@#1\endcsname
\fi
#2}}
\providecommand{\BIBdecl}{\relax}
\BIBdecl

\bibitem{collier2017}
S.~E. Collier, ``The emerging enernet: Convergence of the smart grid with the
  internet of things,'' \emph{IEEE Industry Applications Magazine}, vol.~23,
  no.~2, pp. 12--16, 2017.

\bibitem{palensky2011}
P.~Palensky and D.~Dietrich, ``Demand side management: Demand response,
  intelligent energy systems, and smart loads,'' \emph{IEEE Transactions on
  Industrial Informatics}, vol.~7, no.~3, pp. 381--388, 2011.

\bibitem{singh2017}
S.~Singh, S.~L. Arun, and M.~P. Selvan, ``Regression based approach for
  measurement of current in single-phase smart energy meter,'' in \emph{Proc.
  2017 IEEE Region 10 Symposium (TENSYMP)}, pp. 1--5.

\bibitem{singh2018}
S.~Singh, A.~Roy, and M.~P. Selvan, ``Smart load node for nonsmart load under
  smart grid paradigm: A new home energy management system,'' \emph{IEEE
  Consumer Electronics Magazine}, vol.~8, no.~2, pp. 22--27, March 2019.

\bibitem{giaconi2018}
G.~Giaconi, D.~G{\"u}nd{\"u}z, and H.~V. Poor, ``Smart meter privacy with
  renewable energy and an energy storage device,'' \emph{IEEE Transactions on
  Information Forensics and Security}, vol.~13, no.~1, pp. 129--142, 2018.

\bibitem{zeifman2011}
M.~Zeifman and K.~Roth, ``Nonintrusive appliance load monitoring: Review and
  outlook,'' \emph{IEEE Transactions on Consumer Electronics}, vol.~57, no.~1,
  2011.

\bibitem{molina2010}
A.~Molina-Markham, P.~Shenoy, K.~Fu, E.~Cecchet, and D.~Irwin, ``Private
  memoirs of a smart meter,'' in \emph{Proceedings of the 2nd ACM workshop on
  embedded sensing systems for energy-efficiency in building}.\hskip 1em plus
  0.5em minus 0.4em\relax ACM, 2010, pp. 61--66.

\bibitem{bergman2011}
D.~{Bergman}, D.~{Jin}, J.~{Juen}, N.~{Tanaka}, C.~{Gunter}, and A.~{Wright},
  ``Nonintrusive load-shed verification,'' \emph{IEEE Pervasive Computing},
  vol.~10, no.~1, pp. 49--57, Jan 2011.

\bibitem{shaloudegi2016}
K.~Shaloudegi, A.~Gy{\"o}rgy, C.~Szepesvari, and W.~Xu, ``{SDP} relaxation with
  randomized rounding for energy disaggregation,'' in \emph{Proc. Advances in
  Neural Information Processing Systems}, 2016, pp. 4978--4986.

\bibitem{kumar2017}
A.~Kumar and H.~K. Meena, ``Non-intrusive load monitoring based on graph signal
  processing,'' in \emph{Proc. 2017 Recent Developments in Control, Automation
  \& Power Engineering (RDCAPE)}, pp. 18--21.

\bibitem{ebrahim2018}
A.~F. Ebrahim and O.~Mohammed, ``Household load forecasting based on a
  pre-processing non-intrusive load monitoring techniques,'' in \emph{Proc.
  2018 IEEE Green Technologies Conference (GreenTech)}, April, pp. 107--114.

\bibitem{marchiori2011}
A.~{Marchiori}, D.~{Hakkarinen}, Q.~{Han}, and L.~{Earle}, ``Circuit-level load
  monitoring for household energy management,'' \emph{IEEE Pervasive
  Computing}, vol.~10, no.~1, pp. 40--48, Jan 2011.

\bibitem{lu2017}
T.~{Lu}, Z.~{Xu}, and B.~{Huang}, ``An event-based nonintrusive load monitoring
  approach: Using the simplified viterbi algorithm,'' \emph{IEEE Pervasive
  Computing}, vol.~16, no.~4, pp. 54--61, October 2017.

\bibitem{Liu2019}
Q.~{Liu}, K.~M. {Kamoto}, X.~{Liu}, M.~{Sun}, and N.~{Linge}, ``Low-complexity
  non-intrusive load monitoring using unsupervised learning and generalized
  appliance models,'' \emph{IEEE Transactions on Consumer Electronics},
  vol.~65, no.~1, pp. 28--37, Feb 2019.

\bibitem{singh-SM2019}
S.~Singh and M.~P. Selvan, ``A smart energy meter enabling self-demand response
  of consumers in smart cities of {T}amil {N}adu,'' presented at the
  International Conference on Smart City Model (ICSCM 2019), IIT Madras, India,
  Jan 2019, unpublished.

\bibitem{chapa2007}
\BIBentryALTinterwordspacing
J.~Chapa. (2007, Aug) {THE} {ENERGY} {ORB}: Visualize {E}lectricity
  {C}onsumption!. Accessed: 30 Jan 2019. [Online]. Available:
  \url{https://inhabitat.com/the-energy-orb-monitor-your-electricity-bill/}
\BIBentrySTDinterwordspacing

\bibitem{rich2006}
\BIBentryALTinterwordspacing
S.~Rich. (2006, Feb) {DIY} {KYOTO’S} {WATTSON}. Accessed: 30 Jan 2019.
  [Online]. Available: \url{https://inhabitat.com/diy-kyotos-wattson/}
\BIBentrySTDinterwordspacing

\bibitem{stauffer2012}
\BIBentryALTinterwordspacing
W.~N. Stauffer. (2012, Aug) Tomorrow's power grid. Accessed: 30 Jan 2019.
  [Online]. Available: \url{http://energy.mit.edu/news/tomorrows-power-grid/}
\BIBentrySTDinterwordspacing

\bibitem{ranade2010}
V.~V. Ranade and J.~Beal, ``Distributed control for small customer energy
  demand management,'' in \emph{Proc. 2010 Fourth IEEE International Conference
  on Self-Adaptive and Self-Organizing Systems}, Budapest, pp. 11--20.

\bibitem{tneb2017}
\BIBentryALTinterwordspacing
TANGEDCO. (2017, Aug) Revised tariff rates with effect from 11.08.2017 approved
  rate and payable by the consumer. Accessed: 30 Jan 2019. [Online]. Available:
  \url{https://www.tangedco.gov.in/linkpdf/ONE_PAGE_STATEMENT.pdf}
\BIBentrySTDinterwordspacing

\end{thebibliography}

\end{document}